\documentclass[11pt]{amsart}
\usepackage{amsmath, amssymb, amsfonts}
\usepackage[all]{xypic}
\usepackage[pdftex]{graphicx}
\usepackage{stmaryrd}
\usepackage[flushleft]{paralist} 
\usepackage{todonotes} 
\usepackage{url} 

\usepackage[ruled,vlined]{algorithm2e}

\theoremstyle{plain}
\newtheorem{theorem}{Theorem}[section]
\newtheorem{lemma}[theorem]{Lemma}
\newtheorem{corollary}[theorem]{Corollary}

\theoremstyle{remark}

\newtheorem{remark}[theorem]{Remark}

\newtheorem*{note*}{Note}
\newtheorem*{remark*}{Remark}
\newtheorem*{example*}{Example}

\theoremstyle{definition}
\newtheorem*{definition*}{Definition}
\newtheorem*{hypothesis*}{Hypothesis}
\newtheorem*{assumptions*}{Assumptions}
\newtheorem{definition}[theorem]{Definition}

\newcommand{\Z}{\mathbb{Z}}

\newcommand{\rsa}{\mathbf{RSA}}
\newcommand{\rabin}{\mathbf{Rabin}}
\newcommand{\rsaplus}{\mathbf{RSA+}}
\newcommand{\fact}{\mathbf{Fact}}

\numberwithin{equation}{section}

\title{RSA+: An RSA variant}

\author[S.~Kleine]{Sören Kleine} 
\address[Kleine]{Institut für Anwendungssicherheit, Universität der Bundeswehr München, Wer\-ner-Hei\-sen\-berg-Weg 39, 85577 Neubiberg, Germany} 
\email{soeren.kleine@unibw.de}
\urladdr{https://www.unibw.de/datcom/mitarbeiter/dr-soeren-kleine} 

\author[A.~Nickel]{Andreas Nickel}
\address[Nickel]{Universit\"{a}t der Bundeswehr M\"{u}nchen\\
	Fakult\"{a}t f\"{u}r Informatik\\
	Wer\-ner-Hei\-sen\-berg-Weg 39\\
	85579 Neubiberg\\
	Germany}
\email{andreas.nickel@unibw.de}
\urladdr{https://www.unibw.de/timor/mitarbeiter/univ-prof-dr-andreas-nickel}

\author[T.~Ritter]{Torben Ritter} 
\address[Ritter]{Universität der Bundeswehr München, Wern\-er-\-Hei\-sen\-berg-Weg 39, 85577 Neubiberg, Germany} 
\email{torben.ritter@unibw.de}

\author[K.~Shankar]{Krishnan Shankar} 
\address[Shankar]{Department of Mathematics \& Statistics, James Madison University, 60 Bluestone Drive, Harrisonburg VA 22807} 
\email{shankakx@jmu.edu} 
\urladdr{https://www.jmu.edu/mathstat/people/faculty-full-time/shankar-ravi.shtml} 

\def \br #1 {{\color{blue} #1 }}


\begin{document}

\maketitle

\begin{abstract}
We introduce a new probabilistic public-key cryptosystem which combines the main ingredients of the well-known RSA and Rabin cryptosystems.  We investigate the security and performance of our new scheme in comparison to the other two.
\end{abstract}

\section*{Introduction}

The RSA and Rabin cryptosystems (see \cite{rsa, rabin}) certainly are among the best-known public-key cryptosystems. In both schemes, the computations are done in the unit group of the ring of integers modulo $n$, a composite number with two large prime factors $p$ and $q$. The integer $n$ forms (part of) the public key of each user, and the prime factors $p$ and $q$ are kept secret. 

Whereas the RSA scheme is a widely used public-key cryptosystem, the scheme of Rabin is well-known mainly for theoretical reasons: in contrast to RSA, it is known that an attacker who is able to break the Rabin cryptosystem can factor the public modulus $n$. Therefore the safety of the Rabin cryptosystem relies on the very-well studied hard mathematical problem of factoring a composite number with two large prime factors (the same security level is conjectured to hold true also for RSA, but this has not yet been proven). On the other hand, the Rabin cryptosystem has certain disadvantages from the practical point of view: The decryption function in the Rabin cryptosystem always returns four possible clear texts, and therefore the user is faced with the additional task to decide which clear text is the correct message (but see Section~\ref{section:number-of-solutions} below). 

The motivation for the present note was to somehow combine the RSA and Rabin schemes in order to create, by using elements from the RSA~scheme, a new cryptosystem which on the one hand is more practical than Rabin's cryptosystem and at the same time has a security level which is as close as possible to the security level of the Rabin cryptosystem. A related reason for pursuing RSA+ is that some attacks on RSA use knowledge of the public exponent $e$ to derive $d$, the decryption exponent, and this protocol helps obfuscate the encryption exponent.  We begin by describing our scheme and then investigate the pros and cons of our new approach. More precisely, we study the runtime of our scheme and compare it with plain RSA and Rabin implementations in Section~\ref{section:runtime}. The third section is devoted to a security analysis of the cryptosystem, and in Section~\ref{section:number-of-solutions} we describe an advantage of our scheme over the classical Rabin cryptosystem. 

\section{The cryptosystems} \label{section:cryptosystems}

We begin by briefly recapitulating the (text\-book) RSA and Rabin schemes. Afterwards we introduce our newly proposed cryptosystem. 

\subsection{(Textbook) RSA} 
Let $p$ and $q$ be two large distinct prime numbers of size at least 1500 bits each and let ${n = p \cdot q}$. We assume that ${2p \le q \le 8p}$.\footnote{It is recommended to choose $p$ and $q$ of the same size in RSA, but to keep one of them, say $q$, at least one or two bits longer than $p$ (otherwise the product $N = p \cdot q$ can be factored easily by using an old method which is due to Fermat). One can find different concrete suggestions for the sizes of $p$ and $q$ in the literature. For this paper we consider primes with about $1500-3000$ bits, and we have chosen the condition $2p \leq q \leq 8p$.} Bob's public key is $(n,e)$ where $e$ is coprime with ${\varphi(n) = (p-1)(q-1)}$. His private key is $(p,q,d)$, where $d$ has been chosen such that ${d e \equiv 1 \pmod{\varphi(n)}}$. 

Note that $d$ is efficiently  computable via Euclid's algorithm if $p$ and $q$ are known. In fact, the knowledge of the inverse $d$ of $e$ modulo $\varphi(n)$ is equivalent to the knowledge of the prime factors $p$ and $q$ of $n$. 

\textbf{Encryption.} In order to send a message ${m \in (\Z/n\Z)^\times}$ to Bob, Alice computes 
\[ c \equiv m^e \pmod{n}. \]  

\textbf{Decryption.} Bob can decrypt the message via ${m \equiv c^d \pmod{n}}$. 

\begin{remark} \label{rem:efix} 
  In practical implementations of the RSA scheme one often chooses \emph{first} the encryption exponent ${e = 65537}$ (this is a prime number), and \emph{then} one chooses prime numbers $p$ and $q$ such that $e$ does not divide ${p-1}$ or ${q-1}$. This choice for $e$ is made because encryption with this exponent ${e = 2^{16} + 1}$ is particularly efficient (see \cite[p.~469]{schneier}). 
\end{remark} 

\subsection{Rabin} 
The public key is just ${n = p \cdot q}$, where $p$ and $q$ are as above. 

\textbf{Encryption.} A message ${m \in (\Z/n\Z)^\times}$ is encrypted by computing ${c \equiv m^2 \pmod{n}}$. 

\textbf{Decryption.} First the square-roots of a given ciphertext $c$ (lifted from $(\Z/n\Z)^\times$) modulo $p$ and $q$ are computed. Then the Chinese Remainder Theorem (CRT) is used in order to obtain four different square-roots of $c$ modulo $n$. It remains to decide which of them corresponds to the original message $m$ (see also Section~\ref{section:number-of-solutions} below). 

\subsection{RSA+} 
As above ${n = p \cdot q}$ is the product of two large prime numbers. The public key of Bob is just $n$, as in the Rabin scheme. 

\textbf{Encryption.} If Alice wants to encrypt a message ${m \in (\Z/n\Z)^\times}$, she first finds a random number $x$ coprime with $\varphi(n)$ (for example, one can choose ${x > \sqrt{n}}$ which has passed a strong pseudo-primality test). Then she computes 
\[ c \equiv m^x \pmod{n}\] and 
\[ y \equiv x^2 \pmod{n}\] and transmits the pair ${(c,y)}$ to Bob. 

\textbf{Decryption.} Using his knowledge of $p$ and $q$ Bob computes the four square roots ${x_1, \ldots, x_4}$ of $y$ modulo $n$. For each such square root he tries to compute the inverse \[ u_i \equiv x_i^{-1} \pmod{\varphi(n)}\] 
and in case this is possible he computes 
\[ c^{u_i} \pmod{n}. \] 
The original message $m$ is among the $c^{u_i}$.  

We provide an implementation of the encryption and decryption algorithms in PARI and Python on GitHub.\footnote{see \url{https://github.com/soeren-kleine/RSA_plus-an-RSA-variant}.}  

\begin{remark} \label{rem:x} 
	If $x$ was chosen smaller than $\sqrt{n}$, then ${y = x^2}$ as integers and it would be easy to find the square-root $x$ of $y$ modulo $n$. Therefore it is important to choose ${x > \sqrt{n}}$. There is also another reason for this restriction: The parameter $x$ chosen in the encryption step has to be coprime with $\varphi(n)$ in order to be invertible modulo $\varphi(n)$, i.e. otherwise decryption would not work. However, the person who encrypts a message $m$, using the public key of Bob, does not know ${\varphi(n) = (p-1)(q-1)}$. It is obvious that $x$ has to be odd. However, depending on the choice of $p$ and $q$, the integer $\varphi(n)$ is likely to be divisible also by other small prime factors. One could avoid this by restricting to \emph{safe primes} $p$ and $q$ (i.e. by insisting that both $(p-1)/2$ and $(q-1)/2$ are also prime numbers, which are then so-called Sophie-Germain primes). We don't want to pursue this approach in this note because it considerably restricts the set of possible candidates for $p$ and $q$. 
	
	Instead we use the fact that any \emph{prime number} ${x > \sqrt{n}}$ is automatically coprime with $\varphi(n)$ for sufficiently large $n$ (this is okay if $p$ and $q$ have been chosen to the order of magnitude as described above). Since a primality proof of a large integer is very expensive we instead suggest to use a strong pseudo-prime $x$. If the pseudo-primality test used in this step of the algorithm is good enough then the probability of $x$ being not coprime with $\varphi(n)$ is negligibly small. Note that we will suggest a much more efficient way for choosing an appropriate value for $x$ in Section~\ref{subsection:efficient} below.  
\end{remark} 

\begin{remark} \label{rem:4-roots} 
	In the decryption algorithm of RSA+ the Chinese Remainder theorem gives four different square-roots of $y$ modulo $n$. However, since $n$ is odd, exactly two of them will be even (more precisely, if 
	\[ x_i \in \{1, \ldots, n-1\}\] 
	is odd, then ${n - x_i}$ will be even). Therefore the corresponding $x_i$ cannot be inverted modulo $\varphi(n)$. This shows a first difference with the Rabin cryptosystem: We obtain at most two possible decryptions for each ciphertext $c$, whereas Rabin's decryption algorithm yields 4 different plain texts which have to be checked. 
	
	It is also possible that in fact only one of the four square-roots ${x_1, \ldots, x_4}$ of $y$ modulo $n$ is coprime with $\varphi(n)$ (note that at least one of the four $x_i$ will be the original choice of $x$, and therefore will be coprime to $\varphi(n)$ if $x$ was chosen properly (see also Remark~\ref{rem:x})). We will investigate the number of valid solutions of the decryption algorithm further in Section~\ref{section:number-of-solutions} below. 
\end{remark} 

In the remainder of this section we prove two auxiliary results on the efficient computation of square roots modulo some prime number $p$. These results are well-known, but we include them for the convenience of the reader. 
\begin{lemma} 
\label{lemma:sq1} 
  Let ${p \equiv 3 \pmod{4}}$ be a prime, and let ${y \in (\Z/p\Z)^\times}$ be a quadratic residue. Then the two square roots of $y$ modulo $p$ are 
  \[ x \equiv \pm y^{(p+1)/4} \pmod{p}. \]
\end{lemma} 
\begin{proof} 
   This can be verified by a direct computation, using the fact that 
   \[ y^{(p-1)/2} \equiv 1 \pmod{p}\] 
   because $y$ is a quadratic residue modulo $p$. 
\end{proof} 
\begin{lemma} 
\label{lemma:sq2} 
  Suppose that ${p \equiv 5 \pmod{8}}$, and let ${y \in (\Z/p\Z)^\times}$ be a quadratic residue. 
  
  Let ${\varepsilon = y^{(p-1)/4}}$. Then $\varepsilon$ is either congruent to 1 or $-1$ modulo $p$, and the two square roots of $y$ modulo $p$ are 
  \[ x = \pm \begin{cases} y^{(p+3)/8} & \text{ if $\varepsilon \equiv 1 \pmod{p}$} \\ 
  	                2^{(p-1)/2} \cdot y^{(p+3)/8} & \text{ if $\varepsilon \equiv -1 \pmod{p}$}. \end{cases} \]
\end{lemma} 
\begin{proof} 
	The statement on $\varepsilon$ follows as in the proof of the previous lemma. If ${\varepsilon \equiv 1 \pmod{p}}$, then we let ${x = y^{(p+3)/8}}$ and we compute 
	\[ x^2 \equiv y^{(p+3)/4} = \varepsilon \cdot y \equiv y \pmod{p}. \] 
	If ${\varepsilon \equiv -1 \pmod{p}}$, then the assertion can be proved similarly by using that 
	\[ 2^{(p-1)/2} \equiv -1 \pmod{p}\] 
	because $2$ is a quadratic non-residue modulo $p$ as ${p \equiv 5 \pmod{8}}$. 
\end{proof} 

If ${p \equiv 1 \pmod{8}}$ one can use the Tonelli-Shanks algorithm \cite{tonelli,Shanks} to compute a square root,
but this is significantly more costly.
In view of these facts, we will restrict to primes which are not congruent to $1$ modulo $8$ in all what follows.

\section{Runtime analysis} \label{section:runtime} 
In this section we first analyse the efficiency of our scheme from a theoretical point of view and compare it with the classical RSA and Rabin cryptosystems. 

The public and secret keys in the RSA+ cryptosystem are exactly as in the Rabin cryptosystem (and therefore the key generation is less costly as in the RSA scheme if the encryption exponent $e$ is chosen as in textbook RSA (see Remark~\ref{rem:efix})). An RSA+ ciphertext consists of two integers modulo $N$, i.e. it it twice as long as an RSA or Rabin ciphertext. 

For encryption of a plain text via RSA+, we first have to find a suitable integer $x$ which is coprime with $\varphi(n)$ (this step in theory is as expensive as the choice of the encryption exponent in the public key of the RSA scheme, but in practice it is more difficult because $\varphi(n)$ is not known at this point. Moreover, this step has to be re-done for each encryption). Then we have one RSA encryption step, and one Rabin encryption. 

For the decryption of an RSA+ ciphertext, we first have a Rabin decryption step. Then we have to do one or two modular inversions modulo $N$, followed by one or two RSA decryptions. 

It turns out that the most expensive step is the random choice of $x$ in the RSA+ encryption function. We have made some runtime tests with 2000 bit primes $p$ and $q$, and it turned out that the rather slow search for the random exponent $x$ (including pseudo-primality tests for large integers) makes RSA+ very much slower than textbook RSA (in our tests, it needed more than 40 times more time than RSA). In order to derive a competitive variant of RSA+, we propose the following improved version of our scheme. 

\subsection{A practical implementation} \label{subsection:efficient} 
We describe a more efficient implementation of our RSA+ scheme. The main change will concern the choice of $x$, since it is very expensive to find a large number in the range ${[\sqrt{n}, n]}$ which is a strong pseudoprime. In fact, what we really want is that $x$ is coprime with 
\[ \varphi(n) = (p-1)(q-1).\] 
The main idea is as follows: We choose a random pseudoprime $\ell_0$ which is considerably smaller than $n$, but still large enough to ensure that it is very unlikely that $\ell_0$ divides the unknown integer $\varphi(n)$. Then this base prime $\ell_0$ is multiplied by a power $\ell_1^k$ of a very small random prime $\ell_1$. Here $k$ is chosen such that the product 
\[ x = \ell_0 \cdot \ell_1^k \] 
lies in the interval $[\sqrt{n}, n]$ (recall that we want $x$ to be greater than $\sqrt{n}$ since otherwise it could be easily obtained from the congruence ${y \equiv x^2 \pmod{n}}$). We recommend to choose $\ell_0$ to be a random prime in the range $[2^{150}, 2^{190}]$. There are enough primes in this interval to make an exhaustive search for $x$ very expensive. On the other hand, for integers of this size a probabilistic test for pseudo-primality is still quite fast, so that the choice of $\ell_0$ does not slow down the encryption function of RSA+ too much. Moreover, in our tests we chose $\ell_1$ to be a very small prime (say, up to $100$) which is coprime with $\varphi(n)$. Since $\varphi(n)$ is not publicly known, we added $\ell_1$ to the public key, i.e. it is chosen in the key generation phase of the cryptosystem. Note that the security of the scheme is not weakened by making $\ell_1$ publicly available, since a possible attacker only learns that $p-1$ and $q-1$ are not divisible by $\ell_1$. In the \lq worst case' ${\ell_1 = 3}$ the attacker can exclude half of the prime numbers, but there are enough choices left for $p$ and $q$. 
\begin{algorithm} 
	\DontPrintSemicolon
	\SetKwFunction{Mod}{Mod} \SetKwFunction{Randomprime}{RandomPrime} \SetKwFunction{NextPrime}{NextPrime} \SetKwFunction{Random}{Random} \SetKwFunction{BitLength}{BitLength}
	\SetKwInOut{Input}{Input} \SetKwInOut{Output}{Output}
	
	\SetKwData{SubK}{SubK}
	
	\Input{$n = pq$ is the product of two primes satisfying $4p \le q \le 8p$; \\ a small random prime $l_1$ coprime with $(p-1)(q-1)$; \\ and a message ${m \in (\Z/pq\Z)^\times}$}
	\Output{$(c,y) \in (\Z/pq\Z)^\times \times (\Z/pq\Z)^\times$, an RSA+-encryption of $m$}
	\BlankLine
	$b = \BitLength{p}$ \; 
	$k$ = \Random{$[\lfloor(((b - 148)/\textup{log}_2(x_1))+1)\rfloor, \lfloor((3/2*b-188)/\textup{log}_2(x_1))\rfloor]$} \; 
	\BlankLine 
	$l_0 = \Randomprime{$[2^{150}, 2^{190}]$}$ \; 
    $x = l_0 \cdot l_1^k$ \; 
	\BlankLine 
	$c = \Mod(m,n)^x$ \; 
	$y = \Mod(x,n)^2$ \; 
	\BlankLine 
	\Return{$(c,y)$}
	\caption{Efficient implementation of RSA+ encryption}
	\label{alg:rsa+_efficient}
\end{algorithm}  

We give a schematic overview of this improved encryption function in Algorithm~\ref{alg:rsa+_efficient}. Note that the exponent $k$ in this algorithm is chosen such that $x$ lies between $\sqrt{n}$ and $n$. 

How can we ensure that $x$ as chosen above is coprime with $\varphi(n)$? The very small prime number $\ell_1$ is chosen coprime with $\varphi(n)$ by construction. The moderately small prime $\ell_0$ might be a divisor of $\varphi(n)$, since the person who encrypts a message does not know $\varphi(n)$. However, since $\ell_0$ has at least $150$ bits, the probability of being a divisor of $\varphi(n)$ is negligibly small (in our tests, we did never encounter this exceptional case). 

It turned out that this way of choosing $x$ speeds up the RSA+ functions by more than a factor of ten (for 2000 bit primes $p$ and $q$; the effect is even larger for larger bit sizes). We also implemented an improvement on the decryption function: There are at most two square roots $x_1$ and $x_2$ of $y$ modulo $n$ which are coprime with $\varphi(n)$. We do not consider further the remaining two square roots, i.e. we return only two possible messages instead of four (and in many cases one can indeed sort out one 
of the two remaining messages). Moreover, we use the Chinese Remainder Theorem and our knowledge of the prime factors $p$ and $q$ of $n$ in order to efficiently compute the powers $c^{x_1^{-1}}$ and $c^{x_2^{-1}}$ modulo $n$. This speeds up the decryption moderately. 

\subsection{Runtime comparison} \label{subsection:runtime_comparison} 
In our final runtime tests we generated 1000 keys and for each such key we chose 100 messages which we encrypted and afterwards re-decrypted using our RSA+ - functions, and also via textbook RSA and textbook Rabin. For RSA we used the speed-up from Remark~\ref{rem:efix}. 

The computations were done on a customary laptop with the python programs which we published in our github repository\footnote{ \url{https://github.com/soeren-kleine/RSA_plus-an-RSA-variant}}.  
The results are depicted in the following table. Here \emph{bit length} means the approximate bit length of $p$ and $q$. More precisely, if the bitlength is $b$, then $p$ lies between $2^b$ and $2^{b+1}$, and $q$ lies between $2^{b+2}$ and $2^{b+3}$. For example, if ${b = 2000}$, then ${n = p \cdot q}$ will have between 4002 and 4004 bits. Moreover, we considered only prime numbers $p$ and $q$ which are not congruent to 1 modulo 8, since for such primes $p$ and $q$ particularly efficient routines for computing square roots modulo $p$ and $q$ exist (see Lemmas~\ref{lemma:sq1} and \ref{lemma:sq2} above). In the following table we depict times in milli seconds on average (for one encryption and immediate decryption). 
\begin{center} 
\begin{tabular}{c|c|c|c} 
bit length & RSA+ & RSA & Rabin \\ \hline 
2000 & 33.688 & 12.740 & 9.653 \\ 
1500 & 17.173 & 6.678 & 5.100
\end{tabular} 
\end{center} 
We also did a smaller test with 3000 bit primes (here we considered only 40 pairs of keys, each used for the encryption and decryption of 50 messages). Here are the results: 
\begin{center} 
\begin{tabular}{c|c|c|c} 
	bit length & RSA+ & RSA & Rabin \\ \hline 
	3000 & 91.875 & 35.637 & 26.937 
\end{tabular} 
\end{center} 

We have used many PARI functions in our python code, since typically the PARI routines performed much faster than similar algorithms from other libraries (for example, we compared runtimes with the routines from the sympy library). Therefore we also did runtime tests directly in PARI  (the corresponding PARI source code is also available through the github repository). The results from the PARI runtime tests were in accordance with the above results. 

Summarising, we can conclude that the fast variant of our RSA+ scheme is slower than textbook RSA by a factor of only 2 to 3. 
\begin{remark} 
	In our tests we compared the textbook versions of the three cryptographic procedures. For a real-world application each of the three schemes would have to be enhanced by, for example, some padding routine (one must not encrypt plaintexts directly in order to prevent dictionary attacks). Since these enhancements are more or less the same for all the three schemes, a comparison of the running times of the real-world versions of the three cryptosystems would yield even closer results.
\end{remark}

\section{On the security of the RSA+ cryptosystem} \label{section:security} 
In the last section we have seen that RSA+ is slower than both the RSA and Rabin cryptosystems. In the next two sections we describe advantages of RSA+ over these two schemes. The current section focusses on security issues. We will prove that RSA+ is at least as secure as RSA (see Theorem~\ref{thm:B} below). The ultimate goal would be to prove that the security of RSA+ depends only on the problem of factoring $n$. We argue that in some sense RSA+ seems to be closer to this goal than RSA (we will make this more precise below). First we introduce some terminology. 
\subsection{Security definitions} 
In this subsection we recall, for the convenience of the reader, several notions of different kinds of attacks on a cryptosystem. We start with the notion of a \emph{semantically secure} cryptosystem (see \cite[Section~5.2]{sem_Sicherheit} and \cite[Section~5.1]{sem_Sicherheit2}). This security model typically is defined in terms of a game between a \lq good guy' (let's call him Bob) and an adversary (let's call him Oscar). 
\begin{compactenum}[(1)] 
	\item Oscar chooses two messages $m_0$ and $m_1$ and gives them to Bob. 
	\item Bob randomly chooses a bit ${i \in \{0,1\}}$ and encrypts the message $m_i$. He sends the corresponding ciphertext $c$ back to Oscar. 
	\item Oscar has to decide whether $c$ is the encryption of $m_0$ or $m_1$. 
\end{compactenum} 
The cryptosystem is called semantically secure if Oscar cannot efficiently guess the correct value of $i$ in Step~(3) with a probability which is significantly larger than $1/2$. Here efficiently usually means that Oscar has to make a decision within a time which depends on the size of the parameters of the cryptosystem in a polynomial way (the space need\-ed for his computations usually is also limited). The property of being semantically secure is sometimes also called \emph{ciphertext indistinguishability}, see e.g. \cite[Section~3.2]{Modern_Cryptography}. 

Semantic security describes a notion of security against passive attacks. In order to deal with an active adversary, one has to modify the above game. This leads to stronger attacks as a \emph{chosen-plaintext attack} (CPA) or an \emph{(adaptive) chosen-ciphertext attack} (CCA).

\begin{remark}
	As textbook RSA (see \cite[\S 1.3.5]{Hinek}), the RSA+ cryptosystem in its pure form is not semantically secure. Though it is probabilistic in nature (unlike textbook RSA!),
	the Jacobi symbol $\left(\frac{m}{n}\right)$ of a plain text $m$ is revealed by the cyphertext $(c,y)$. It is indeed straightforward
	to see that one has $\left(\frac{m}{n}\right) = \left(\frac{c}{n}\right)$, since the encryption exponents $x$ used in RSA+ are all odd integers. As explained in \cite[\S 12.5]{Ferguson-Schneier}, a solution to avoid this problem is to use only squares as plain texts.
\end{remark}

\subsection{On the security of RSA, Rabin and RSA+} 
In this subsection, we let $n = p \cdot q$ be as in Section~\ref{section:cryptosystems}. We first formulate a bunch of problems and then prove relations between these problems. 
\begin{definition} Let $n$ be as above, and let ${c,y \in (\Z/n\Z)^\times}$. 
	\begin{compactenum}[(1)] 
		\item[$(\mathbf{Fact}(n))$] Construct a black box which can compute the two prime factors $p$ and $q$ of $n$. 
		\item[$(\rsa(n))$] Construct a black box that can decrypt \emph{any} RSA encrypted message modulo $n$. 
		\item[$(\rsa(c,n))$] Construct a black box that can decrypt the given RSA encrypted message ${c \in (\Z/n\Z)^\times}$. 
		\item[$(\rabin(n))$] Construct a black box which can decrypt \emph{any} Rabin encrypted message modulo $n$. 
		\item[$(\rabin(c,n))$] Construct a black box which can decrypt the given Rabin encrypted message ${c \in (\Z/n\Z)^\times}$. 
		\item[$(\rsaplus(n))$] Construct a black box that can decrypt \emph{any} RSA+ encrypted message modulo $n$. 
		\item[$(\rsaplus(c,y,n))$] Construct a black box that can decrypt the given RSA+ encrypted message $(c,y)$. 
	\end{compactenum} 
\end{definition} 
For example, by a black box for problem $(RSA+(n))$, we mean
that there is some kind of unknown method or algorithm that takes as input any given tuple $(c,y)$, 
where $c = m^x \pmod n$ and $y = x^2 \pmod n$ as above, and returns $m$ and potentially a second
unintelligible message. Note that it is unreasonable to expect that the black box always returns only $m$
as this would mean that it can distinguish between intelligible and unintelligible messages. Similary, a black box for $(Rabin(n))$ will output, for any given ${c \in (\Z/n\Z)^\times}$, the four square roots of $c$ modulo $n$ (but it cannot decide which of the square roots corresponds to the original message $m$). 

First note that there are some obvious relations between the above problems. For example, if one can solve $(\rsa(n))$ for $n$, then the corresponding black box can of course solve $(\rsa(c,n))$ for any ${c \in (\Z/n\Z)^\times}$. In the following we will write such a relation as 
\[ (\rsa(n)) \Longrightarrow (\rsa(c,n)) \; \forall c \in (\Z/n\Z)^\times.\] 
Similarly, ${(\rabin(n)) \Longrightarrow (\rabin(c,n))}$ and ${(\rsaplus(n)) \Longrightarrow (\rsaplus(c,y,n))}$ for all \\ ${c,y \in (\Z/n\Z)^\times}$. 

More importantly, we have the following well-known
\begin{theorem} 
  $(\rabin(n)) \Longleftrightarrow (\rabin(c,n))$ $\Longleftrightarrow (\fact(n)) \Longrightarrow (\rsa(n))$ for each $c, n$ as above. In other words, breaking Rabin is equivalent to factoring $n$, and factoring $n$ is enough for breaking RSA. 
\end{theorem} 
\begin{proof} 
  It is obvious that a black box which can compute the prime factors $p$ and $q$ of $n$ can solve both problems $(\rsa(n))$ and $(\rabin(n))$. On the other hand, it is well-known that any black box which can solve $(\rabin(c,n))$ for some $c$ can factor $n$. For the convenience of the reader we briefly recapitulate the argument. Suppose that we input $(c,n)$ into the black box and receive the four square roots ${x_1, x_2, x_3, x_4}$ of $c$ modulo $n$. These square roots come in pairs; without loss of generality we can assume that 
  \[ x_2 \equiv -x_1 \pmod{n}, \quad x_4 \equiv -x_3 \pmod{n} \quad \text{ and } x_1 \not\equiv \pm x_3 \pmod{n}.\] 
  
Since these are all square roots of the same number we have
 \begin{align*} (x_1 + x_3) \cdot (x_1 - x_3) & = x_1^2 - x_3^2 \\ 
  	& \equiv c - c = 0 \pmod{n}, \end{align*}
so it follows that $n \mid (x_1 + x_3)(x_1 - x_3)$, but $n\nmid (x_1 + x_3)$ nor $n\nmid (x_1 - x_3)$.  But the primes $p,q$ both divide $n$, so $p,q$ both divide the product $(x_1 + x_3)(x_1 - x_3)$ and hence they each divide one of the factors.  If they were to both divide the same factor, say $x_1 - x_3$, then it would follow that $n = pq\mid (x_1 - x_3)$ which is not allowed.  Thus  $p \mid (x_1 + x_3)$ and ${q \mid (x_1 - x_3)}$, or vice versa. Since the $x_i$ can be computed by the black box algorithm, we have factored $n$. 
  
  Finally, the chain of implications 
  	\[ (\rabin(c,n)) \Longrightarrow (\fact(n)) \Longrightarrow (\rabin(n)) \] 
  	shows that the two problems $(\rabin(n))$ and $(\rabin(c,n))$ are in fact equivalent for any choice of ${c \in (\Z/n\Z)^\times}$.  
\end{proof} 
\begin{corollary} 
	$(\rabin(n)) \Longrightarrow (\rsa(n))$, i.e. the Rabin cryptosystem is at least as secure as RSA. 
\end{corollary} 
It is \emph{conjectured}, but not known, that in fact 
\[ \rsa(n) \Longleftrightarrow (\fact(n)), \] 
which would also imply that breaking $(\rsa(n))$ and $(\rabin(n))$ is equivalent. 

Now we turn to RSA+. In the following result we assume that there is a black box that can decrypt RSA+ messages. We prove that the same black box also solves the problem $(\rsa(n))$. This means that breaking RSA+ is at least as difficult as breaking the RSA cryptosystem, i.e. in some sense RSA+ is at least as secure as RSA. 

\begin{theorem} \label{thm:B} 
	For any $n$ we have $(\rsaplus(n)) \Longrightarrow (\rsa(n))$. 
\end{theorem} 
\begin{proof}
	Suppose we have a black box that is able to decrypt RSA+ messages as described above.
	Now, if $(n,e)$ is an RSA public key and ${c \equiv m^e \pmod n}$ is a given RSA ciphertext,
	the input $(c,e^2)$ into the black box returns $m$ and potentially a second possible message $m'$. Even if $m'$ were a meaningful message,
	we could single out $m$ by checking if ${c \equiv m^e \pmod n}$ or ${c \equiv (m')^e \pmod{n}}$.
\end{proof}

\begin{remark}
The above represents the worst case scenario i.e., we are assuming that the black box
can decipher all RSA+ ciphertexts and then it follows that any RSA ciphertext can also be
deciphered. We will show -- under reasonable hypotheses -- that also the average complexity of RSA+ is at least as good as that of RSA.
See Remark \ref{rem:average-complexity} below.
\end{remark}

\begin{remark}
If $\rsa(n)$ is equivalent to factoring $n$, then the same holds true for $\rsaplus(n)$. Though we cannot prove the equivalence of $\rsaplus(n)$ and $\fact(n)$, Theorem \ref{thm:B} suggests that $\rsaplus(n)$ is somewhat `closer' to $\fact(n)$ than $\rsa(n)$. Note that most (if not all) known attacks on RSA make use of the public exponent $e$. If one tries to apply such an attack to
RSA+, however, one would have to compute the exponent $x$ first, for which one would have to break Rabin's cryptosystem;
but the latter is known to be equivalent to $\fact(n)$. The same is true if the attacker manages to find an inverse $\tilde{x}$ to $x$ modulo $\varphi(n)$, by which he could easily compute the clear text $m$ from a ciphertext $c$. But then $\tilde{x}^2$ was the inverse of ${y = x^2}$ modulo $\varphi(n)$, and it is well-known that the knowledge of both $y$ and its inverse modulo $\varphi(n)$ allows for efficient factorization of $n$.
\end{remark}

\begin{remark} \label{rem:twomessages} 
Let $(c,y)$ be an RSA+ ciphertext. It is quite common that the decryption procedure yields two possible plaintext messages (see also the next section for more details on how likely this result will happen). If an attacker is given two tuples $(m_1, x_1)$ and $(m_2, x_2)$ such that ${m_1^{x_1} \equiv c \equiv m_2^{x_2} \pmod{n}}$ and such that both $x_1$ and $x_2$ are square roots of $y$ modulo $n$, then he could easily factor $n$. This is because the two square-roots of $y$ must satisfy ${x_1 \not\equiv -x_2 \pmod{n}}$ since $-x_1$ will be even and thus will not be coprime with $\varphi(n)$. 
	   
It is not clear to us whether the knowledge of two potential plaintext messages $m_1$ and $m_2$ without knowing the corresponding encryption exponents $x_1$ and $x_2$ would suffice for factoring $n$. If this was true, then breaking RSA+ was equivalent to factoring $n$, since receiving two possible plaintext messages can always (see the next section) be ensured via suitable choice of the parameters. 
\end{remark} 

\section{On the number of possible clear texts output by the decryption function} \label{section:number-of-solutions} 

A major drawback of Rabin's cryptosystem is the fact that decryption leads to four possible messages. If the original message 
was text, it is usually easy to make the right choice. However, if the message was a number, this is much more difficult.
So it is an advantage of RSA+ over Rabin that it only leads to at most two possible messages. Of course, this can also be achieved for Rabin's cryptosystem by simply adding the parity of the message to the cyphertext. This then always leads to two possible messages.

If both primes dividing $n$ are congruent to $3 \pmod 4$, then a workaround for Rabin has been suggested by Williams
\cite{williams}. The same method works for RSA+: If Bob chooses $x$ to be a square mod $n$, then only one of the four square roots of $y = x^2 \pmod n$, namely $x$ itself, is a square mod $n$. This follows from the fact that $-1$ is not a quadratic residue
mod $p$ nor mod $q$. The RSA+ decryption procedure can be modified to ignore the second odd square root of ${y \pmod{n}}$ and proceed with ${u \equiv x^{-1} \pmod{n}}$. Note that breaking this modified Rabin scheme is no longer equivalent to factoring $n$; an analogous remark can be made for RSA+ (see also Remark~\ref{rem:twomessages}). 

However, even if one of the primes dividing $n$ is congruent to $1 \pmod 4$ or if $x$ is not chosen to be a square mod $n$,
the RSA+ decryption scheme sometimes only produces one possible message. This happens whenever the second odd square root
of $y \pmod n$ is not coprime with $\varphi(n)$. The probability of a randomly chosen integer to be divisible by a prime
$\ell$ is $1/\ell$. Let us denote the set of odd primes dividing $\varphi(n)$ by $\mathcal{L}_n$. Then heuristically, the
possibility that decryption only produces one possible message is given by the formula

\begin{equation} \label{eqn:heurisitc-formula}
	1 - \prod_{\ell \in \mathcal{L}_n} \left(1 - \frac{1}{\ell}\right). 
\end{equation} 

For instance, if $\varphi(n)$ is divisible by $3$, $5$ and $7$, then this probability is at least 
\[ 1-\frac{2}{3}\cdot \frac{4}{5}\cdot \frac{6}{7} = \frac{57}{105} > \frac{1}{2}.\] 
So we expect that there is only one possible message at least in every second case. In a series of tests we observed an accuracy of formula \eqref{eqn:heurisitc-formula} of roughly $99\%$. We used pairs $(p,q)$ of primes of bit length $1500$ and for each of a few hundred pairs of such primes we RSA+-encrypted and decrypted $10^4$ messages. Then we checked whether decryption yields one or two possible results.

In other words, by choosing $n$ appropriately one can enlarge the possibility that decryption returns only one message.  
In particular small prime factors of $\varphi(n)$ have relatively large impact. Of course, if all prime factors of either
$p-1$ or $q-1$ are small, this makes the cryptosystem vulnerable to factoring methods such as Pollard's $p-1$ method
\cite{pollard}. 

On the other hand, there do always exist choices for $x$ which result in two possible plaintext messages returned by the decryption procedure. Indeed, it suffices to exclude two possible residue classes modulo each prime divisor ${\ell \in \mathcal{L}_n}$ (more precisely, we want $x$ and $n-x$ to be not divisible by $\ell$, i.e. we exclude the residue classes of 0 and $n$ modulo $\ell$). This means that by choosing random values for $x$ we have a good chance to find exponents which are not contained in any of the \lq forbidden' residue classes (recall that the person who encrypts a message does not know the prime factors $p$ and $q$). Note that two possible plaintext decryptions of the same ciphertext might be used for factorization of $n$; see Remark~\ref{rem:twomessages}. 

\begin{remark} \label{rem:average-complexity}
	We discuss the average complexity of RSA+ compared to that of RSA.
	Fix a modulus $n$. Let $\mathcal{C}$ be the set of all pairs $(c,e)$ where $e$ is a valid RSA exponent and $c = m^e$
	is a corresponding RSA ciphertext. Similarly, we let $\mathcal{C}^+$ be the set of all valid RSA+ ciphertexts $(c,y)$,
	that is $c = m^e$ as before and $y=e^2$. Then we have a surjective map
	\begin{eqnarray*}
		\mathrm{plus}: \mathcal{C} & \rightarrow & \mathcal{C}^+\\
		(c,e) & \mapsto & (c,e^2).
	\end{eqnarray*}
	As explained above, for a given $(c,y) \in \mathcal{C}^+$ there are one or two preimages under this map. If ${a \in [0,1]}$ denotes the 
	probability that there are two preimages, then one has $|\mathcal{C}| = (1+a)|\mathcal{C}^+|$.
	Now suppose that there is a black box that can decipher $b\%$ of all RSA+ ciphertexts. Let $U^+ \subseteq \mathcal{C}^+$
	be the set of all ciphertexts that the black box can decipher, and let $U = \mathrm{plus}^{-1}(U^+)$ be its preimage.
	It follows as in the proof of Theorem \ref{thm:B} that the black box can
	decipher an RSA ciphertext $c$ whenever $(c,e) \in U$. 
	If we assume that $U^+$ is uniformly distributed in $\mathcal{C}^+$, then the probability of having two preimages for a random $(c,y) \in U^+$ is still $a$.
	Then
	\[
		\frac{|U|}{|\mathcal{C}|} = \frac{(1+a)|U^+|}{(1+a)|\mathcal{C}^+|} = \frac{|U^+|}{|\mathcal{C}^+|} = \frac{b}{100},
	\]
	i.e.\ the black box can also decipher $b\%$ of all possible RSA messages for the given modulus $n$.
	Of course, there are also two extreme scenarios: (i) the black box can only decipher $(c,y) \in \mathcal{C}^+$
	that possess only one preimage, and (ii) it can only decipher $(c,y)$ that possess two preimages. In case (i)
	the black box can only decipher $(1+a)^{-1}b\%$ of all possible RSA messages, whereas in case (ii)
	it can decipher $2(1+a)^{-1}b\%$. Note that the latter is strictly greater than $b\%$ unless $a=1$ (in which case
	both $p$ and $q$ must be Fermat primes). 
\end{remark}

\section{Conclusion} 
In this article we proposed a novel public-key crypto system which somehow combines the encryption methods of the well-known RSA and Rabin cryptosystems. We have seen that the runtime of our algorithm, if implemented efficiently, is not much more than twice the runtime of the RSA cryptosystem. On the other hand, breaking our cryptosystem is at least as difficult as breaking the RSA scheme, and it seems reasonable to expect that breaking our new algorithm is equivalent to factoring the large modulus $n$, which is widely believed to be a very hard mathematical problem. When compared to the original Rabin scheme, our algorithm has the benefit of producing at most two (compared to four) possible decrypted messages from a fixed ciphertext -- in many situations we obtain in fact a unique decrypted plaintext.

\section{Acknowledgments}
The fourth named author would like to thank Kimball Martin, Marco Streng and Larry Washington for useful comments on an early draft of the manuscript.

\bibliography{references}{}

\providecommand{\bysame}{\leavevmode\hbox to3em{\hrulefill}\thinspace}
\providecommand{\MR}{\relax\ifhmode\unskip\space\fi MR }
\providecommand{\MRhref}[2]{%
  \href{http://www.ams.org/mathscinet-getitem?mr=#1}{#2}
}
\providecommand{\href}[2]{#2}
\begin{thebibliography}{RSA78}

\bibitem[FS03]{Ferguson-Schneier}
N.~Ferguson and B.~Schneier, \emph{Practical cryptography}, Hoboken, NJ: John
  Wiley \& Sons, 2003 (English).

\bibitem[GM84]{sem_Sicherheit}
S.~Goldwasser and S.~Micali, \emph{Probabilistic encryption}, J. Comput. System
  Sci. \textbf{28} (1984), no.~2, 270--299. \MR{760548}

\bibitem[Hin10]{Hinek}
M.~J. Hinek, \emph{Cryptanalysis of {RSA} and its variants}, Chapman \&
  Hall/CRC Cryptography and Network Security, CRC Press, Boca Raton, FL, 2010.
  \MR{2554564}

\bibitem[KL08]{Modern_Cryptography}
J.~Katz and Y.~Lindell, \emph{Introduction to modern cryptography}, Chapman \&
  Hall/CRC Cryptography and Network Security, Chapman \& Hall/CRC, Boca Raton,
  FL, 2008. \MR{2371431}

\bibitem[Pol74]{pollard}
J.~M. Pollard, \emph{Theorems on factorization and primality testing}, Proc.
  Cambridge Philos. Soc. \textbf{76} (1974), 521--528. \MR{354514}

\bibitem[Rab79]{rabin}
M.~O. Rabin, \emph{Digitalized signatures and public-key functions as
  intractable as factorization}, MIT Laboratory for Computer Science
  \textbf{MIT-LCS-TR 212} (1979).

\bibitem[RSA78]{rsa}
R.~L. Rivest, A.~Shamir, and L.~Adleman, \emph{A method for obtaining digital
  signatures and public-key cryptosystems}, Commun. ACM \textbf{21} (1978),
  120--126 (English).

\bibitem[Sch15]{schneier}
B.~Schneier, \emph{Applied cryptography. {Protocols}, algorithms and source
  code in {C}. 20th anniversary edition. {With} a foreword by {Whitfield}
  {Diffie}}, reprint of the 1996 2nd ed., Hoboken, NJ: John Wiley \& Sons, 2015
  (English).

\bibitem[Sha73]{Shanks}
D.~Shanks, \emph{Five number-theoretic algorithms}, Proceedings of the {S}econd
  {M}anitoba {C}onference on {N}umerical {M}athematics ({U}niv. {M}anitoba,
  {W}innipeg, {M}an., 1972), Congress. Numer., vol. No. VII, Utilitas Math.,
  Winnipeg, MB, 1973, pp.~51--70.

\bibitem[Sho98]{sem_Sicherheit2}
V.~Shoup, \emph{Why chosen ciphertext security matters}, IBM Research Report RZ
  3076, IBM Research Division, 1998.

\bibitem[Ton91]{tonelli}
A.~Tonelli, \emph{Bemerkung über die {A}uflösung quadratischer
  {C}ongruenzen}, Nachrichten von der Königl. Gesellschaft der Wissenschaften
  und der Georg-Augusts-Universität zu Göttingen \textbf{1891} (1891),
  344--346.

\bibitem[Wil80]{williams}
H.~C. Williams, \emph{A modification of the {RSA} public-key encryption
  procedure}, IEEE Trans. Inf. Theory \textbf{26} (1980), 726--729 (English).

\end{thebibliography}
\bibliographystyle{amsalpha}

\end{document}